\documentclass[reprint,
superscriptaddress,
amsmath,amssymb,
aps,
pra,
]{revtex4-2}

\usepackage{graphicx}
\usepackage{dcolumn}
\usepackage{bm}
\usepackage{xcolor}
\usepackage[percent]{overpic}
\usepackage{color}
\usepackage[table]{xcolor}
\usepackage{booktabs}
\usepackage[hidelinks]{hyperref}
\usepackage{colortbl}
\usepackage{braket}
\usepackage{gensymb}
\hypersetup{
    colorlinks,
    linkcolor={red!50!black},
    citecolor={blue!50!black},
    urlcolor={blue!80!black}
}
\usepackage{hyperref}

\makeatletter
\providecommand\href@noop[2]{#2}
\makeatother

\begin{document}

\setlength{\textfloatsep}{10pt}

\preprint{APS/123-QED}

\title{Helically Enhanced Chiroptical Response and Symmetry Breaking in Conjugated Polymers }

\author{\small Aaron Forde}
\email{aforde@lanl.gov}
\affiliation{\small Theoretical Division, Los Alamos National Laboratory, Los Alamos, NM, 87545, U.S.A.}

\author{\small Braden M. Weight}
\affiliation{\small Theoretical Division, Los Alamos National Laboratory, Los Alamos, NM, 87545, U.S.A.}

\author{\small Prashanna Poudel}
\affiliation{\small Department of Physics and Astronomy, University of Utah, Salt Lake City, Utah 84112, U.S.A }

\author{\small Avadh Saxena}
\affiliation{\small Theoretical Division, Los Alamos National Laboratory, Los Alamos, NM, 87545, U.S.A.}
\affiliation{\small Center for Nonlinear Studies, Los Alamos National Laboratory, Los Alamos, NM, 87545, U.S.A.}

\author{\small Zeev Valy Vardeny}
\affiliation{\small Department of Physics and Astronomy, University of Utah, Salt Lake City, Utah 84112, U.S.A }

\author{\small Christoph Boehme}
\affiliation{\small Department of Physics and Astronomy, University of Utah, Salt Lake City, Utah 84112, U.S.A }

\author{\small Alan Bishop}
\affiliation{\small Theoretical Division, Los Alamos National Laboratory, Los Alamos, NM, 87545, U.S.A.}
\affiliation{\small Center for Nonlinear Studies, Los Alamos National Laboratory, Los Alamos, NM, 87545, U.S.A.}

\author{\small Sergei Tretiak}
\email{serg@lanl.gov}
\affiliation{\small Theoretical Division, Los Alamos National Laboratory, Los Alamos, NM, 87545, U.S.A.}
\affiliation{\small Center for Integrated Nanotechnologies, Los Alamos National Laboratory, Los Alamos, NM, 87545, U.S.A.}

\date{\today}

\begin{abstract}
Chiral $\pi$-conjugated polymers are an attractive material platform for spin polarized carrier-transport and spectroscopy, but fundamental considerations for how torsional disorder influences the response properties of the material have not been considered. Here we combine atomistic electronic structure modeling with with experimental spectroscopic measurements to examine symmetry breaking in the prototypical $\pi$-conjugated polymer polyacetylene, (CH)$_x$. Chiral (CH)$_x$ oligomers are generated in distinct conformations which differ in their out-of-plane tonsorial ordering. We find that a \textit{helical }conformation introduces orders of magnitude enhanced chiroptical activity due to a solenoid effect. This effect is visualized by the Transition Chiral Tensor analysis which shows signatures of domain ordering which eliminates destructive interference between electric and magnetic contributions. These findings highlight the capability to develop a hierarchical interpretation relating local, fragment symmetry breaking to global, nonlocal interactions governing chiroptical response in emerging chiral materials. 

\end{abstract}

\maketitle

{\small
\textit{Introduction -} Organic $\pi$-conjugated polymers have provided an alternative to inorganic crystalline materials for electronic and opto-electronic devices. Their developments for technological applications have necessitated an understanding of their fundamental physics leading to developments in theoretical models, computational methods, and ultra-fast spectroscopy.  Strong electron-phonon coupling gives rise to non-linear excitations in the form of solitons\cite{Su1979} (for degenerate ground-states), polarons\cite{Fesser1983} (for degenerate and non-degenerate ground-states), and the dynamical breather modes\cite{Bishop1984,Phillpot1989,Tretiak2003}. Electronic correlations have been found to have a dramatic effect on the relative ordering of dipole-forbidden and dipole-allowed electronic excited-states $2A_g$ and $1B_u$\cite{Soos1993}, respectively, with exciton-phonon interactions describing the localization of excitations\cite{Sergei2002prl}.  Initial theoretical and computational models relied on semi-empirical approaches, such as the SSH\cite{Su1979}, or PPP\cite{Soos1993} models, but are now accompanied by a variety of atomistic quantum chemistry methods ranging from semiempirical methods\cite{Tretiak2002} to density-functional theory\cite{Keya2022} (DFT) and its time-dependent extension (TD-DFT)\cite{Lin2006}, and to multi-reference method such as DMRG\cite{Chan2015}, and many-body based GW+BSE\cite{Tiago2004}.
}

{\small
The impact of chiral symmetry breaking on charge-carrier transport, namely the chirality induced spin-selectivity (CISS) effect \cite{Ray2006,Xie2011} as observed in insulating DNA oligomers, has motivated synthesizing chiral $\pi$-conjugated polymers. In addition to the impact on carrier transport  chiral symmetry breaking endows the preferential absorption and emission of circularly polarized light (CPL). Inducing CPL sensitivity into $\pi$-conjugated polymers offers the opportunity to develop novel photodiode devices\cite{Furlan2024}, such as photodetectors \cite{chen2026} and circularly polarized  light sources\cite{Tiffany2025}. Methods to synthesize chiral $\pi$-conjugated polymers have included liquid crystals \cite{Akagi1998},  chiral dopants \cite{Li2019,Wan2022}  asymmetric catalysis \cite{Han2018}, and mechanical shearing\cite{Sun2019}. The earliest report of helical polyacetylene (CH)$_x$\cite{Akagi1998} manifested as fibrils with an estimated torsional dihedral angle between adjacent monomers ranging from $0.02\degree$ to $0.23\degree$. Later chirality induction has been extended to donor-acceptor block polymers which have applications in opto-electronics, such as F8BT\cite{Wan2023} and PII2T\cite{Sun2024}.
}


{\small
Despite all of the exciting developments, a theoretical understanding of chiral $\pi$-conjugated polymers is far from complete. In the context of the optical response, a predictive scheme that quantitatively links the degree of chirality in the spatial structure, electronic structure, and optical response is desirable. For example,  $\pi$-conjugated polyenes have been interpreted to be most stable as planar oligomer chains due to their $sp^2$ hybridized bonding\cite{ZHU1992}. Polyenes are members of point groups containing 2-fold rotation axes, mirror symmetries, and inversion symmetries with irreducible representations, such as $D_{2h}$ for \textit{cis} and $C_{2v}$ for \textit{trans} conformers, which provide optical selection rules. Chiral symmetry breaking necessitates removing the mirror and inversion symmetries allowing for chiroptical activity manifested as a mirrored optical rotation for chiral enantiomers. However, details linking the structural details of the chiral structure, such as torsion angle and helical pitch, to the observables governing chiroptical response, such as the relative orientation of electric and magnetic transition dipoles, $\vec{d}^{0\alpha}$ and $\vec{m}^{0\alpha}$, respectively, are necessary. Establishing these structure-to-property relationships for chiral $\pi$-conjugated materials will help guide synthetic chemists aim for a desired target with their products. 

To this end, we perform atomistic \textit{ab initio} electronic structure calculations to investigate the prototypical $\pi$-conjugated polymer \textit{cis}-(CH)$_x$ and the changes to its electronic structure and optical response with chiral symmetry breaking. Chiral \textit{cis}-(CH)$_x$ oligomers are produced via \textit{twist} and \textit{helix} conformations shown in \textbf{Figure 1}. In both the electronic structure and optical response the models show contrasting behavior. In the optical response the \textit{twist} model retains identical optical selection rules as the $D_{2h}$ symmetry group whereas in the \textit{helical} model symmetry-forbidden dark states become bright. From a fragment decomposed symmetry analysis of the transition electric dipoles we find the origin to be out-of-plane contributions mixing even and odd parities. In the chiroptical response the \textit{helical} structure shows two orders of magnitude increased intensity compared to the \textit{twist} due to a solenoid effect increasing the magnitudes of the transition magnetic dipole. We spatially resolve the site contribution of monomer fragments to the observed rotatory strength using the Transition Chiral Tensor analysis which clearly visualizes domain ordering in the \textit{helix} which eliminates destructive interference between the electronic and magnetic transition densities. These results suggest a hierarchical interpretation linking details in the local structural chirality to the observed global chiroptical response.   
 }

\begin{figure}[t!]
    \centering
    \includegraphics[width=\columnwidth,clip,trim = 0 0 0 0]{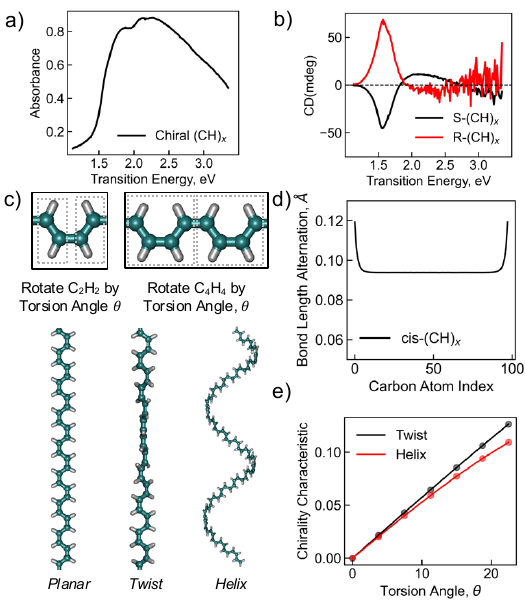}
    \caption{ \footnotesize  
    (a) Optical absorption and (b) circular dichroism of chiral (CH)$_x$ films. (c) Chiral cis-(CH)$_{x}$ oligomers are generated from the linear D$_{2h}$ symmetry chain to make the \textit{twist} and \textit{helix} chains. (d) Bond-Length Alternation $r_n$ = (-1)$^{n}$ ($d_{n-1} - d_{n}$) demonstrating the dimerization. Elongated bonds at the boundaries are a signature of the free-boundary conditions. (e) Chirality characteristic (Eq 1) of the \textit{twist} and \textit{helix} oligomers as a function of torsion angle demonstrating the breaking of mirror and inversion symmetries.  
    }
    \label{FIG:STRUCTURE}            
\end{figure}
{\small
\textit{Chiral Oligomer:} We begin by demonstrating the existence of chiral symmetry breaking in films of CH$_x$ with details of the synthesis and spectroscopy provided in the \textbf{Supplemental Material}. \textbf{Figure 1}\textbf{a} shows the absorption spectrum of the unannealed film which contains a distribution of \textit{trans}-(CH)$_x$ and \textit{cis}-(CH)$_x$ where the absorption onset is at $1.5$ $eV$. \textbf{Figure 1}\textbf{b} shows the associated circular dichroism (CD) spectra for the films prepared with an enantiomer chiral precursor. The existence of chiral symmetry breaking is clearly seen by the mirrored CD spectra for the S-(CH)$_x$ and R-(CH)$_x$ films. From microscopy we do not find evidence of helical fibrils on the mesoscale indicating that the chiral symmetry breaking occurs on the microscale. 

For the atomistic modeling we begin with the linear \textit{cis}-(CH)$_{x}$ model oligomer of D$_{2h}$ symmetry and apply torsion to generate out-of-plane rotation resulting in broken inversion and mirror symmetries. \textbf{Figure 1}\textbf{c} shows the geometric structures for two distinct types of applied torsion which we refer to as \textit{twist} and \textit{helix} models corresponding to S-enantiomers. \textit{Twisted} structures are generated by sequential dihedral rotation of the entire oligomer by angle $\theta$ at each C$_{2}$H$_{2}$ monomer fragment along the chain where the torsion is applied to the \reflectbox{'}short' \textbf{C-C} bonds. The \textit{helix} was generated in the same manner, but between sequential C$_{4}$H$_{4}$ fragments which make up the \textit{cis}-(CH)$_{x}$ primitive unit cell where torsion is applied to the \reflectbox{'}long' \textbf{C-C} bonds. \textbf{Figure 1}\textbf{d} shows the bond-length alternation (BLA) parameter defined as $r_n$ = (-1)$^{n}$ ($d_{n-1} - d_{n}$) where $n$ is the index of the carbon atom along the oligomer. This represents the dimerization of the chain into alternating \reflectbox{'}short' and \reflectbox{'}long' bonds. The bulk of the chain shows a constant value of $0.93$ $\AA$ which agrees well with experimental\cite{ZHU1992,Yannoni1983} and computational studies including increased correlation \cite{Peralta2024,Windom2022} compared to the range-seperated hybrid  DFT functional used here. To quantify the degree of symmetry breaking in our \textit{twisted} and \textit{helical} \textit{cis}-(CH)$_{x}$ models we employ the chirality characteristic\cite{Ethan2024,Ethan2024prl}
\begin{equation}\label{CC}
    \chi = \sum_{n=1}^{N-2} \frac{ \vec{v}_{n} \cdot (\vec{v}_{n+1} \times \vec{v}_{n+2}) } {|\vec{v}_{n}|~|\vec{v}_{n+1}|~| \vec{v}_{n+2}| },
\end{equation}
where $\vec{v}_n = \vec{r}_{n-1} - \vec{r}_{n}$ and $\vec{r}_{n}$ are the Cartesian coordinates of the $n_\mathrm{th}$ carbon atom along the chain and $N$ is the total number of carbon atoms. In short, this pseudo-vector metric captures the local torsional symmetry breaking and sums the local contributions across the model. The chirality characteristic $\chi$ for the \textit{twisted} and \textit{helical} models for torsion angles ranging from up to $22\degree$ are shown in \textbf{Figure 1}\textbf{e}. This clearly demonstrates that the magnitude of symmetry breaking increases in proportion to the applied torsion and that we are dealing with chiral \textit{cis}-(CH)$_{x}$ oligomers.  
}

\begin{figure}[t]
    \centering
    \includegraphics[width=\columnwidth]{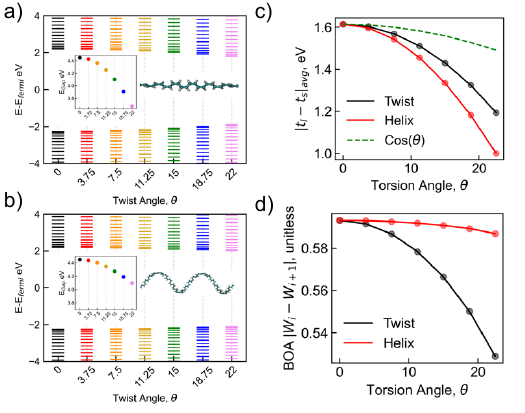}
    \caption{ \footnotesize  
    Density of electronic states for (a) \textit{twist} and (b) \textit{helix} models of \textit{cis}-(CH)$_x$ as a function of torsion angle with the insets showing the bandgap. (c) Change in the difference of average nearest neighbor hopping parameter $|t_l -t_s|_{avg}$ for \reflectbox{'}long' and \reflectbox{'}short' bonds, respectively, over applied torsion for \textit{twist} (red) and \textit{helix} (black) models. The green dashed line is a simple $cos(\theta)$ decay over the applied torsion. (d) Change in BOA over applied torsion angles. 
    }
    \label{FIG:SINGLE_PARTICLE}            
\end{figure} 
{\small
\textit{Torsion Perturbed Electronic Structure:} Next, we examine changes to the electronic properties of the ground state using DFT and optical responses using TD-DFT of the chiral \textit{cis}-(CH)$_{x}$ oligomers. In \textbf{Figure 2} \textbf{a} and \textbf{b}, we show the change in the electronic bandgap for the \textit{twisted} and \textit{helical} oligomers, respectively, by presenting the energies of the single-particle states (\textit{i.e.}, Kohn-Sham orbitals). In both cases, \textit{twisted} and \textit{helical}, a monotonic decrease of the bandgap, shown as insets for each figure, is observed relative to the planar D$_{2h}$ \textit{cis}-(CH)$_{x}$ oligomer. This provides an additional source of spectral broadening (\textbf{Figure 1}a), complementing the commonly accepted broadening arising from the distribution of conjugation lengths. From a simple nearest neighbor (NN) tight-binding model the bandgap is directly proportional to the difference in hopping amplitudes between \reflectbox{'}short' and long' bonds $E_{gap} \propto |t_{s} - t_{l}|$ where \textit{s} and \textit{l} refer to short and long bonds, respectively. Since we keep BLA fixed for each model, the change in the hopping amplitudes is directly attributable to the applied torsion. We estimate the nearest-neighbor hopping parameters from the Fock matrix elements of the Natural Atomic Orbitals (NAOs). NAOs $\phi_i$ are determined from the diagonalization of the density matrix constructed from nonorthogonal  atom-centered AO basis functions \cite{Reed1985}.  Diagonal and off-diagonal Fock matrix elements $t_{ij} = \braket{\phi_i|\hat{F}|\phi_j}$ correspond to site and hopping energies, respectively. \textbf{Figure 2}\textbf{c} displays the change in averaged nearest-neighbor hopping parameters $|t_l - t_s|_{avg}$ as a function of applied torsion. We observe a monotonic decrease of the averaged difference for both the \textit{twist} and \textit{helix} models which is anticipated from the observed decrease in the bandgaps. Interestingly, the rate of decrease for both models is faster than a simple $cos(\theta)$ dependence on the torsion angle $\theta$, shown as the green-dotted line in \textbf{Figure 2}\textbf{c}. This indicates a reduced electronic coupling with the increased torsion. In contrast, the useful chemical concept of bond-order (i.e. the number of electrons shared in a chemical bond) is appropriate for $\pi$-conjugated polymers to quantify the change in electron distribution with the applied torsion. Wiberg bond orders\cite{Wiberg1968} are determined from the density matrix $P_{ij}$ (in NAO basis) between long or short \textbf{C-C} fragments $W_{AB} = \sum_{i\in A} \sum_{j\in B} (P_{ij})^2$. \textbf{Figure 2}\textbf{d} shows the bond-order alternation (BOA) $|W_i - W_{i+1}|$ along the oligomer as function of applied torsion. Here we observe a more contrasting trend where the \textit{twist} model shows an order of magnitude larger decrease in the BOA than the helix. This indicates a propensity for the \textit{twist} model to break the dimerization.  Interestingly, the BOA rather than the nearest-neighbor hopping paramters, provides a more consistent trend to the electronic structure where the \textit{twist} model shows a faster decrease in the electronic bandgap suggestive of increasing conjugation by decreasing the dimerization character towards metallic behavior. 
}
\begin{figure*}[t!]
    \centering
    \includegraphics[width=\linewidth]{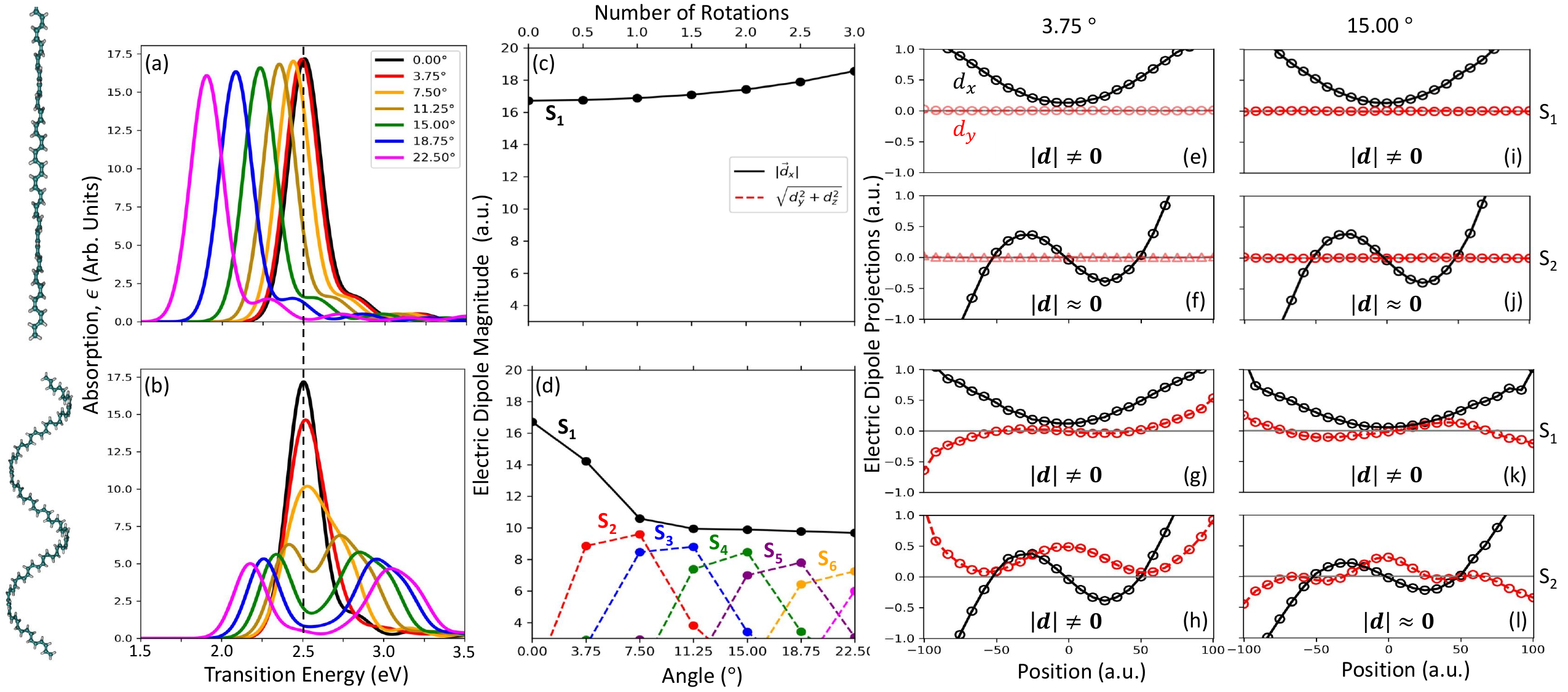}
    \caption{ \footnotesize  
   (a)-(b) \textit{cis}-(CH)$_{x}$ absorption spectrum as function of applied torsion for (a) \textit{twist} and (b) \textit{helical } symmetry breaking. (c)-(d) Analysis of transition dipole moment components for lowest excited-states. (e)-(h) Projection of the transition dipole moments onto molecular fragments along the oligomer arc-length for 3.75$\degree$ torsion. (i)-(l) same as (e)-(h) for 15$\degree$ torsion 
    }
    
    \label{FIG:ABS_DIPOLE}
\end{figure*}

\begin{figure}[t!]
    \centering
    \includegraphics[width=\linewidth,clip,trim = 0 0 0 0]{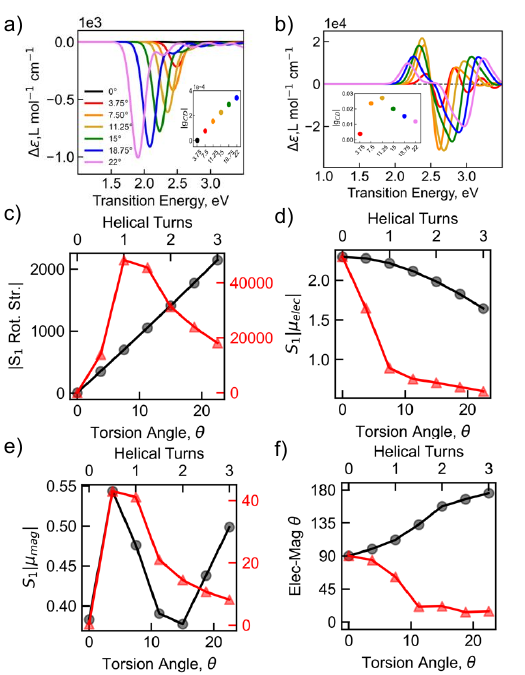}
    \caption{ \footnotesize  
   (a)-(f) Analysis of \textit{S}$_{1}$ optical transition features for \textit{twisted }(black, circles) and \textit{helical }(red, triangles) cis-(CH)$_x$ Change in the (a) optical gap, (b) oscillator strength, (c) rotatory strength, (d) electric transition dipole magnitude (e) magnetic transition dipole magnitude, and (f) angle between electric and magnetic transition dipole with applied torsion (bottom) and the equivalent number of helical turns (top). For (c) and (e) the value for \textit{helical} cis-(CH)$_x$ corresponds to the secondary axis. 
    }
    \label{FIG:ROT_COMPONENTS}           
\end{figure}

\begin{figure}[t!]
    \centering
    \includegraphics[width=\linewidth]{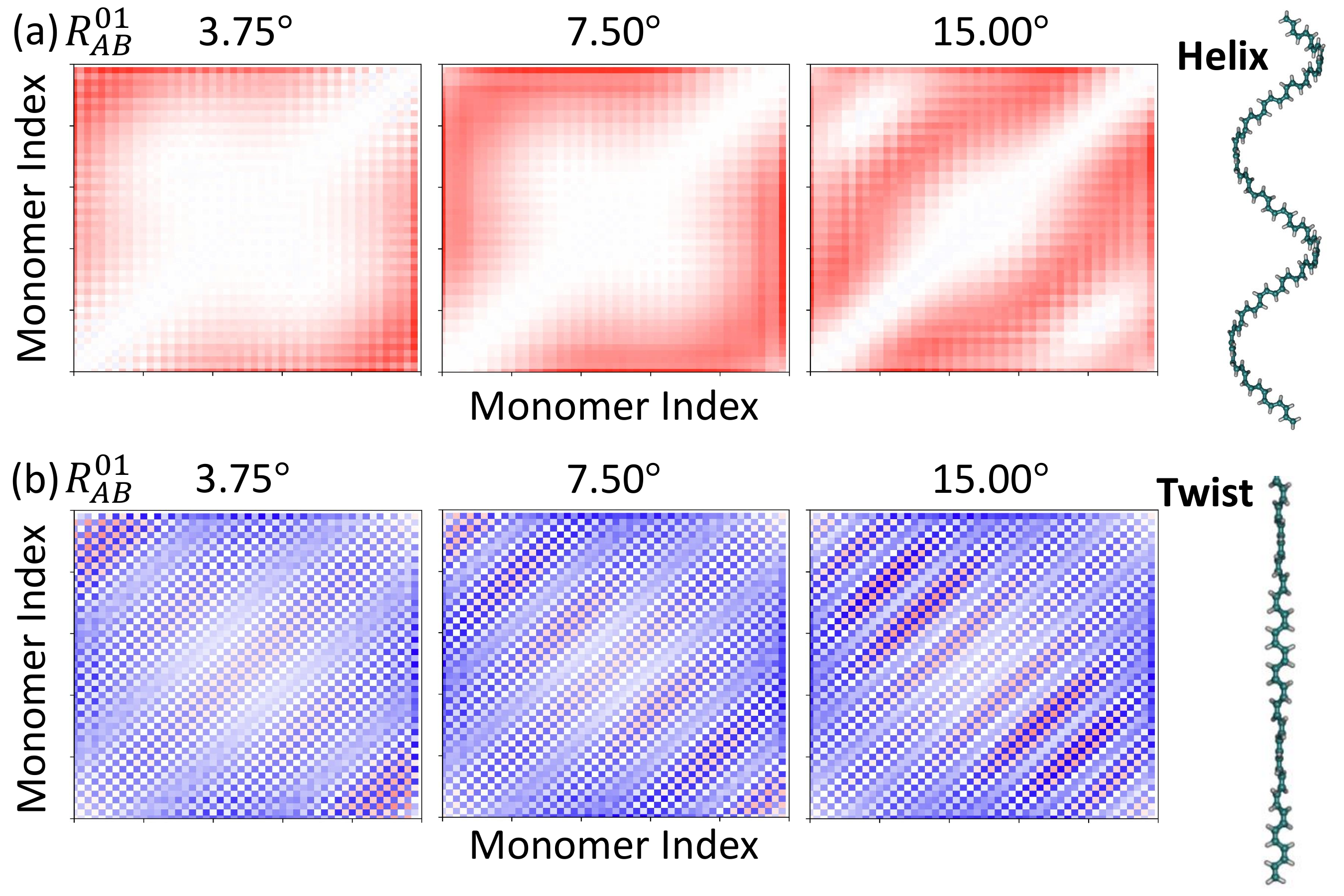}
    \caption{ \footnotesize  
    Transition chiral tensors $\boldsymbol{R_{AB}^{0\alpha}}$ for the (a) helix and (b) twist \textit{cis}-(CH)$_x$ oligomers at applied torsion angles of 3.75$\degree$, 7.50$\degree$, and 15.00$\degree$. In each panel the \textit{x} and \textit{y} axes correspond to the electronic coordinate $\textbf{Q}^{0\alpha}$ and momentum $\textbf{P}^{0\alpha}$ projected  to a $C_2H_2$ monomer along the \textit{cis}-(CH)$_x$ oligomer with the amplitudes demonstrating the non-local interference resulting in positive (red) or negative (blue) contributions to the net optical rotation. 
    }
    \label{FIG:CHIRAL_TENSOR}           
\end{figure}



\small{
\textit{Helical Electric Dipole Symmetry Breaking:} From the electronic structure analysis we next consider spectroscopic features. \textbf{Figure 3}\textbf{a} plots the optical absorption expressed as extinction $\epsilon$ for the \textit{twisted} oligomer. For the planar D$_{2h}$ \textit{cis}-(CH)$_{x}$ it is well established that the lowest single electron excitation is of 1A$_\mathrm{g}$ $\rightarrow$ 1B$_\mathrm{u}$ character (which we will refer to as $S_1$). This excitonic transition is optically bright. In the standing wave picture\cite{Wu2006}, the higher-energy odd excitonic transitions $S_n$ $(n =3,5,7,...)$ show reduced intensities whereas their even order counterparts $(n=2,4,6,...)$ are symmetry forbidden. The computed absorption spectrum for the \textit{twisted} oligomer exactly follows the symmetry described above only modulated by a red-shift in proportion to the torsional angle. The \textit{helical} \textit{cis}-(CH)$_{x}$ displays a contrasting behavior in its optical response as shown in \textbf{Figure} \textbf{3}\textbf{b}. Namely, in the optical absorption we again observe the red-shift of the $S_1$ excitonic transition with increased torsion. However, unlike the \textit{twist} model, the \textit{helix} conformation exhibits a pronounced redistribution of oscillator strength arising from its three-dimensional helical geometry, which modifies the electronic selection rules.
}

\small{
\textbf{Figures 3}\textbf{c} and \textbf{d} compare the components of the transition dipole moment parallel and perpendicular to the polymer axis for the \textit{twist} and \textit{helix} models. These correspond to the $\hat{x}$ (solid lines) and $\hat{y}$ (dashed lines) components, respectively. For the \textit{twist} model when considering the $S_1$ transition only the parallel $\hat{x}$ component contributes to the total moment. This is exactly the same as the planar oligomer of D$_{2h}$ where excited-states form standing waves along a one-dimensional chain. When introducing perpendicular contributions in the \textit{helix} model, the selection rules break down. With increased torsion the parallel $\hat{x}$ component from the $S_1$ state decreases linearly until reaching one full pitch which then remains constant with increased torsion. Concurrently the perpendicular $\hat{y}$ component for $S_2$ increases rapidly up to forming one full helical turn, and then decreases where a series of higher energy $S_n$ states become the \reflectbox{'}bright' state with increased torsion.  


}


\small{
To explicitly demonstrate these effects originate from symmetry considerations,\cite{Frexias2025} we compute the transition dipole moment density and project it onto individual $C_2H_2$ unit cell fragments along the oligomer, as described in the \textbf{End Matter}. \textbf{Figures 3}\textbf{e} and \textbf{f}  show the fragment projected contribution to the dipole moment components of the $S_1$ and $S_2$ state, respectively, for the \textit{twist} model at dihedral rotation of 3.75$\degree$. For the $S_1$ and $S_2$  states we find the parallel $\hat{x}$ components possess even and odd symmetry along the oligomer arc-length, respectively. With increased torsion to 15$\degree$ as shown in \textbf{Figures 3}\textbf{g} and \textbf{h} the symmetry of the moments in each excited-state remain unchanged. When examining the \textit{helix} model, shown in \textbf{Figures 3}\textbf{i-l} for the same excited-state roots and torsion angles, the symmetries of the parallel $\hat{x}$ components remain the same. However, the perpendicular $\hat{y}$ components gain non-vanishing intensity with the both $S_1$ and $S_2$ states showing odd and even symmetry, respectively. This local decomposition analysis demonstrates that the out-of-plane torsional distortion modifies the selection rules by mixing even and odd parity states. This finding has important experimental implications, as linearly polarized spectroscopy can directly probe the direction-dependent components of the dielectric tensor, $\epsilon_{ij}$, enabling direct detection of out-of-plane symmetry breaking in oriented films of chiral polymers. 
}


\small{
\textit{Helically Enhanced Chiroptical Activity:} We next examine the chiroptical response of the models expressed as the differential exctinction coefficent $\Delta \epsilon$ shown in \textbf{Figures 4}\textbf{a} and \textbf{b} for the \textit{twist} and \textit{helix} models, respectively. The \textit{twist} model  exhibits a mono-signate behavior (\textit{i.e.}, no Cotton effect) with an increasing intensity of the rotatory strength in response to an increase in applied torsion.  Notably the overall profile of both spectra is largely retained, even as the twisting angle is increased. The inset of \textbf{Figure 4}\textbf{a} shows the $S_1$ anisotropy factor $g_{CD}$ which ranges from $1\times10^{-4}$ to $4\times10^{-4}$.
In contrast, the \textit{helix} model displays bi-signate behavior, with adjacent positive and negative intensities. Focusing on the $S_1$ excitonic transition, there is a non-linear increase in the intensity from $3.75\degree$ to $11.25\degree$ of torsion and then a continuously decreasing intensity with further increased torsion. This is clearly visualized by examining $S_1$ anisotropy factor $g_\mathrm{CD}$ (which follows the same trend), shown as the inset of \textbf{Figure 4}\textbf{b}. Interestingly, the maximal anisotropy factor $g_\mathrm{CD}$ for the \textit{helix} model is $3\times10^{-2}$, which is 2 orders of magnitude larger than the maximal value in the \textit{twist} model, which is $4\times10^{-4}$. We note that the $S_1$ exciton is delocalized across the oligomer for both the \textit{twist} and \textit{helix} models, as observed by the atom-resolved transition density matrix $\boldsymbol{Q}^{0\alpha}$ in Figures S1 and S2 in the \textbf{Supporting Material}, indicating that changes in the optical response are directly attributable to the strain-induced torsion on the collective electronic response. 
}

{\small
To further delineate the optical response behavior of the $S_1$ excitonic transition from the chiral \textit{cis-}(CH)$_{x}$ models we directly compare the relevant observables involved in the spectral response. The rotatory strength of the $S_1$ state for the \textit{twist} model (\textbf{Figure 4}\textbf{c}) increases nearly linearly with increasing torsion. In contrast, the \textit{helix} exhibits a strongly nonmonotonic dependence, reaching a maximum at approximately one complete helical turn before decreasing with further torsion. The rotatory strength matrix element can be decomposed into its primary components $R^{0\alpha} = \mathrm{Im}(\vec{d}^{0\alpha} \cdot \vec{m}^{\alpha 0})$ where $\vec{d}^{0\alpha} = -\langle 0 | \sum_{i} \vec{\hat{r}}_i | \alpha \rangle$ and $\vec{m}^{\alpha 0} = \langle 0 | \sum_{i} \vec{\hat{r}}_i \times \vec{\hat{p}}_i | \alpha \rangle$ are the electric and magnetic transition dipoles, respectively, and $i$ labels the electrons. \textbf{Figures 3}\textbf{d}, \textbf{e}, and \textbf{f} show the change in the magnitude of the magnetic transition dipole, electric transition dipole, and the angle between them, respectively, for the $S_1$ excited-state. In the planar D$_{2h}$ \textit{cis}-(CH)$_{x}$ configuration at zero torsion, the transition dipoles are orthogonal to one another as anticipated due to the lack of inversion symmetry breaking. With finite torsion, the the symmetry is reduced, breaking orthogonality between the electric and magnetic transition dipole moments and allowing chiroptical activity. At large torsion (up to $22.5\degree$), in both models, the transition dipoles become nearly collinear. In both models, the transition electric dipoles closely mirror the behavior of the oscillator strengths. The transition magnetic dipole, however, exhibit a more complex dependence on molecular conformation. For \textit{twist} models the magnitude remains within $20\%$ of the achiral D$_{2h}$ \textit{cis}-(CH)$_{x}$ oligomer; whereas, for the \textit{helix} geometry, the magnitude of the magnetic dipole increases by two orders of magnitude and then non-monotonically decreases with increased torsion beyond one full twist. This can be interpreted as a solenoid effect where the magnetic flux through the coil is proportional to the surface area of the helix, $\Phi(\theta) \propto \pi r^2(\theta)$.\cite{weight2026}  Since we keep the oligomer arc length fixed, the effective radius $r$ of the helix decreases with increased torsion through $r(\theta) \propto \frac{1}{tan(\theta)}$ where the magnetic flux then decays as $\Phi(\theta) \propto \frac{1}{tan^{2}(\theta)}$. This is a crucial insight into the chiroptical activity of organic semiconductors:  chiral symmetry breaking in both the \textit{twist} and \textit{helix} configurations enables optical activity by making the transition dipoles nearly collinear. However, the magnitude of the chiroptical response can vary by orders of magnitude, following intuitive principles of classical electromagnetism. }

{\small 
The rotatory strength of the $S_1$ excitonic state can be further into decomposed into Transition Chiral Tensors (TCT)\cite{weight2026,Freixas2025chemsci,weightinfluence2026} which describe the non-local interactions of the electronic and magnetic transition dipoles. Expanding into transition dipole densities, as described in the \textbf{End Matter}, we obtain fragment projections of the rotatory tensor $R_{AB}^{0\alpha}$. The matrices $R_{AB}^{0\alpha}$ are visualized in \textbf{Figure 5}, where red and blue indicate positive and negative optical rotation, respectively, implying constructive or deconstructive interactions between fragments. Starting with the \textit{helix} model, \textbf{Figure 5}\textbf{a} shows the TCT for torsional angles 3.75$\degree$, 7.50$\degree$, and 15.00$\degree$ corresponding to $1/2$, $1$, and $2$ full rotations, respectively. In each case there are coherent domains of positive optical rotation contributions (red features) from the local $C_2H_2$ fragments with nodal planes (white features) that increase with increased torsion. In contrast, for the \textit{twist} model in \textbf{Figure 2}\textbf{b}, there are regions of positive and negative optical rotation (red and blue features), which signifies alternating domains providing descriptive interference. Overall, these findings show that the \textit{helical} geometry offers a synergistic enhancement of optical activity by both amplifying the transition magnetic dipole through a solenoid-like effect and supporting coherent domains of optical rotation that eliminate destructive interference.     
}

{\small
\textit{In conclusion}, we have investigated how chiral symmetry breaking fundamentally reshapes the electronic structure and optical response of the prototypical $\pi-$conjugated polymer polyacetylene. Asymmetric catalysis was used to synthesize chiral (CH)$_x$ films that exhibit strong chiroptical behavior,  confirming the formation of chiral enantiomers. First principles electronic structure and optical response calculations were employed to elucidate the hierarchical interplay of chiral symmetry breaking, molecular geometry, electronic structure, and optical response. Model chiral oligomers were generated by applying torsion to \textbf{C-C} bonds along a \textit{cis}-(CH)$_x$ chain in various conformations, producing distinct conformations that differ in the degree of out-of-plane symmetry breaking. While chiral conformers have similar local symmetry breaking, as quantified by the chirality characteristic measure, their electronic and optical responses differ markedly due to the distinct global chiral organization. For the electronic structure, we find a systematic red-shift of the bandgap with increasing torsion, which can be rationalized by a reduction of the nearest-neighbor hopping matrix elements controlling the $\pi$-conjugated electrons. For the optical response, symmetry analysis demonstrates that helical conformations with significant out-of-plane distortions break the optical selection rules of the D$_{2h}$ space group, resulting in the redistribution of oscillator strength. This implies that linearly polarized spectroscopy can be used to probe the direction-dependent components of the dielectric
tensor $\epsilon_{ij}$ for helical symmetry breaking in oriented films. Finally, the \textit{helix} conformer exhibits a chiroptical response that is nearly two orders of magnitude stronger than that of the other conformation. This enhancement arises from a solenoid-like effect that amplifies the transition magnetic dipole through coherent constructive interference of local electronic and magnetic domains along the oligomer. The insights from this work is general for the class of $\pi-$conjugated semiconductors where the chemical complexity may increase for the wider class of systems, but the the role of torsional disorder and the global symmetry breaking are fundamental considerations which extend beyond the polyacetylene material explored here. Overall, our results establish a hierarchical picture in which local chiral symmetry breaking propagates through the electronic structure to determine the global chiroptical response. This framework provides structure-property relationships for the emerging class of chiral $\pi-$conjugated semiconductors. 

}

\vspace{0.25cm}
{\bf \large Acknowledgments}
{\small
B.M.W. appreciates the support of the Director's Postdoctoral Fellowship at Los Alamos National Laboratory (LANL), which is funded by the Laboratory Directed Research and Development (LDRD) at LANL. S.T. acknowledges support of the U.S. Department of Energy, Office of Basic Energy Sciences, Division of Chemical Sciences, Geosciences, and Biosciences under LANLE3T1 work-package. The work at the University of Utah was supported by the NSF grant DMR-2206653. LANL is operated by Triad National Security, LLC, for the National Nuclear Security Administration of the US Department of Energy (Contract No. 89233218CNA000001). Approved for unlimited release (LA-UR-26-25110). The research was performed, in part, at the Center for Integrated Nanotechnologies (CINT), a U.S. Department of Energy, Office of Science user facility at LANL. Computing resources were provided by the LANL Institutional Computing (IC) Program.
}

\bibliographystyle{apsrev4-2}
\bibliography{main.bib}

\clearpage

\section{END MATTER}

Here we describe the methods used to decompose the computed transition density matricies in the molecular orbital basis into fragment projected transition dipoles and the chiral transition tensor.  Excited electronic states were found using time-dependent density functional theory (TDDFT) by solving the Casida equations. Briefly, solving the random phase approximaton (RPA),
\begin{equation}
    \begin{bmatrix}
        \textbf{A}&\textbf{B}\\
        -\textbf{B}&-\textbf{A}
    \end{bmatrix}
    \begin{bmatrix}
        \textbf{X}^{0\alpha}\\
        \textbf{Y}^{0\alpha}
    \end{bmatrix} =
    E^{0\alpha}
    \begin{bmatrix}
        \textbf{X}^{0\alpha}\\
        \textbf{Y}^{0\alpha}
    \end{bmatrix},
\end{equation}
one obtains the transition density matrix
\begin{equation}
    \boldsymbol{\xi}^{0\alpha} = 
    \begin{bmatrix}
    \boldsymbol{0}& \boldsymbol{X}^{0\alpha}\\
    \boldsymbol{Y}^{0\alpha}& \boldsymbol{0}
    \end{bmatrix},~
    {\boldsymbol{\xi}^{0\alpha}}^\dagger = 
    \begin{bmatrix}
    \boldsymbol{0}& \boldsymbol{Y}^{0\alpha}\\
    \boldsymbol{X}^{0\alpha}& \boldsymbol{0}
    \end{bmatrix},
\end{equation}
with normalization $\mathrm{Tr}[\boldsymbol{\xi}^{0\alpha}(\boldsymbol{O}-\boldsymbol{V}){\boldsymbol{\xi}^{0\alpha}}^\dagger] = \textbf{X}^{0\alpha}\cdot \textbf{X}^{0\alpha} - \textbf{Y}^{0\alpha}\cdot \textbf{Y}^{0\alpha}$, where $\boldsymbol{O}$ ($\boldsymbol{V}$) is the identity matrix for the occupied (virtual) MO subspace.

In the collective electronic oscillator (CEO) picture\cite{Tretiak2002}, the electronic coordinate (symmetric) and momentum (anti-symmetric) are written as
\begin{equation}
    \textbf{Q}^{0\alpha} = \frac{1}{\sqrt{2}}\big( \boldsymbol{\xi}^{0\alpha} + {\boldsymbol{\xi}^{0\alpha}}^\dagger\big),
\end{equation}

and
\begin{equation}
    \textbf{P}^{0\alpha} = \frac{1}{\sqrt{2}}\big( \boldsymbol{\xi}^{0\alpha} - {\boldsymbol{\xi}^{0\alpha}}^\dagger\big),
\end{equation}
respectively.

The rotary strength between the ground and excited RPA state can again be decomposed into electric and magnetic components as
\begin{align}\label{EQ:EXCITONIC_EL_DIPOLE}
    \vec{d}^{0\alpha} &= \langle 0 | \vec{\hat{r}} | \alpha \rangle = \mathrm{Tr}[\vec{\boldsymbol{d}}~\textbf{Q}^{0\alpha}]\nonumber\\
    &= \sum_{ia} \vec{d}_{ia} Q^{0\alpha}_{ia},
\end{align}
and
\begin{align}\label{EQ:EXCITONIC_MAG_DIPOLE}
    \vec{m}^{0\alpha} &= \langle 0 | \vec{\hat{m}} | \alpha \rangle= \mathrm{Tr}[\vec{\boldsymbol{m}}~\textbf{P}^{0\alpha}]\nonumber\\
    &= \sum_{ia} \vec{m}_{ia} P^{0\alpha}_{ia},
\end{align}
respectively, with occupied orbital $i$ and virtual orbital $a$. The MO integrals for the electric and magnetic dipole moments are obtained by contraction over the atomic orbital (AO) basis, $\{\mu,\nu\}$. The AO integrals $\vec{d}_{\mu\nu} = \langle \mu | \vec{\hat{d}} | \nu\rangle$ and $\vec{m}_{\mu\nu} = \langle \mu | \vec{\hat{m}} | \nu\rangle$, MO expansion coefficients $c_{\mu i}$, and the transition densities $\xi^{0\alpha}_{ia}$ for each transition $0\rightarrow\alpha$ are directly obtained by the Gaussian16 software package.\cite{g16} Here, $\hat{d} = -\sum_{i} \vec{\hat{r}}_i + \sum_I Z_I \vec{R}_I \mathbb{I}$ is the electric dipole operator, summing over each electron $i$ and nucleus $I$, and $\hat{m} = \sum_{i} \vec{\hat{r}}_i \times \vec{\hat{\nabla}}_i$ is the magnetic dipole operator with $\vec{\hat{\nabla}} = i\vec{\hat{p}}$ is the usual gradient operator. The corresponding rotary strength in the excitonic basis $\{\alpha\}$ is then written as
\begin{equation}
    R^{0\alpha} = \vec{d}^{0\alpha} \cdot \vec{m}^{\alpha 0}.
\end{equation}

The rotary strength in the excitonic basis is then visualized according to Ref.~\citenum{weight2026}. The transition chiral tensor is constructed as
\begin{equation}
    R_{\mu\nu} = \vec{d}^{0\alpha}_{\mu\mu} \cdot \vec{m}^{0\alpha}_{\nu\nu},
\end{equation}
where $\vec{d}^{0\alpha}_{\mu\mu} = \langle \mu | \boldsymbol{\vec{d}}~\boldsymbol{Q}^{0\alpha}|\mu\rangle$ and $\vec{\boldsymbol{m}}^{0\alpha}_{\mu\mu} = \langle \mu | \vec{\boldsymbol{m}}~\boldsymbol{P}^{0\alpha}|\mu\rangle$ are the components of the trace in Eqs.~\ref{EQ:EXCITONIC_EL_DIPOLE} and~\ref{EQ:EXCITONIC_MAG_DIPOLE} in the AO basis. This matrix is then symmetrized as $(R^{0\alpha}_{\mu\nu} + R^{0\alpha}_{\mu\nu})/2 \rightarrow R^{0\alpha}_{\mu\nu}$. Note that the asymmetric part, $(R^{0\alpha}_{\mu\nu} - R^{0\alpha}_{\mu\nu})/2 \rightarrow R^{0\alpha}_{\mu\nu}$, does not contribute to the total rotary strength
\begin{equation}\label{EQ:R_from_Rmn}
    R^{0\alpha} = \sum_{\mu\nu} R^{0\alpha}_{\mu\nu}.
\end{equation}
The $\boldsymbol{Q}^{0\alpha}$ and $\boldsymbol{P}^{0\alpha}$ can be transformed into the AO basis as
\begin{equation}\label{EQ:Q_mo_to_ao}
    Q_{\mu\nu}^{0\alpha} = \sum_{ia} c_{\mu i}~Q^{0\alpha}_{ia}~c_{\nu a}
\end{equation}
and
\begin{equation}\label{EQ:P_mo_to_ao}
    P_{\mu\nu}^{0\alpha} = \sum_{ia} c_{\mu i}~P^{0\alpha}_{ia}~c_{\nu a}.
\end{equation}
Finally, the transition chiral tensors are atom-localized by summing over the AOs localized to each atom or fragment $\{A,B\}$ as 
\begin{equation}\label{EQ:CONDENSED_R_TENSOR}
    R^{0\alpha}_{AB} = \sum_{\mu \in A} \sum_{\mu \in B} R^{0\alpha}_{\mu\nu}.
\end{equation}
Note that the rotatory strength is conserved $R^{0\alpha} = \sum_{\mu,\nu}R_{\mu\nu}^{0\alpha} = \sum_{A,B} R_{AB}^{0\alpha}$.


\end{document}


\preprint{APS/123-QED}

\title{--Supplemental Material -- \\ Helically Enhanced Chiroptical Response and Symmetry Breaking in Conjugated Polymers }

\author{\small Aaron Forde}
\email{aforde@lanl.gov}
\affiliation{\small Theoretical Division, Los Alamos National Laboratory, Los Alamos, NM, 87545, U.S.A.}

\author{\small Braden M. Weight}
\affiliation{\small Theoretical Division, Los Alamos National Laboratory, Los Alamos, NM, 87545, U.S.A.}

\author{\small Prashanna Poudel}
\affiliation{\small Department of Physics and Astronomy, University of Utah, Salt Lake City, Utah 84112, U.S.A }

\author{\small Avadh Saxena}
\affiliation{\small Theoretical Division, Los Alamos National Laboratory, Los Alamos, NM, 87545, U.S.A.}
\affiliation{\small Center for Nonlinear Studies, Los Alamos National Laboratory, Los Alamos, NM, 87545, U.S.A.}

\author{\small Zeev Valy Vardeny}
\affiliation{\small Department of Physics and Astronomy, University of Utah, Salt Lake City, Utah 84112, U.S.A }

\author{\small Christoph Boehme}
\affiliation{\small Department of Physics and Astronomy, University of Utah, Salt Lake City, Utah 84112, U.S.A }

\author{\small Alan Bishop}
\affiliation{\small Theoretical Division, Los Alamos National Laboratory, Los Alamos, NM, 87545, U.S.A.}
\affiliation{\small Center for Nonlinear Studies, Los Alamos National Laboratory, Los Alamos, NM, 87545, U.S.A.}

\author{\small Sergei Tretiak}
\email{serg@lanl.gov}
\affiliation{\small Theoretical Division, Los Alamos National Laboratory, Los Alamos, NM, 87545, U.S.A.}
\affiliation{\small Center for Integrated Nanotechnologies, Los Alamos National Laboratory, Los Alamos, NM, 87545, U.S.A.}

\date{\today}

\maketitle
\tableofcontents
\newpage

\section{Synthesis}

\textit{Materials:} All chemicals were obtained from commercial sources and used as received unless 
otherwise specified and stored in a nitrogen filled dry box. Solvents were of spectrograde or higher and thoroughly degassed via sparging followed by freeze-pump-thaw prior to use and stored in a nitrogen-filled glovebox. Acetylene (electronics grade, 99.95 percent dissolved in DMF) was purchased from Linde and was passed through a Parker Balston AA gas purifier filter prior to deposition. All chemical reactions were conducted in a nitrogen filled dry box unless otherwise specified.

\textit{General method for the synthesis of (BINOLate)Ti(OR)\textsubscript{2} complexes:} To an oven-dried flask is added BINOL (1.0 g, 3.49 mmol, 1 equiv) and a magnetic stir bar under nitrogen. The vessel is sealed with a septum and 20 mL of dry benzene is added. The solution is stirred and Ti(O\textit{\textsuperscript{i}}Pr)\textsubscript{4} (1.05 mL, 3.5 mmol, 1.05 equiv) is added. A reflux condenser was attached and the red-orange solution was stirred for 1h at reflux under nitrogen and then evacuated to dryness. The resulting solution was titrated with 250 $\mu$L of pentane cooled to -30 ℃, and solvent removed via pipet. The resulting solids were evacuated to dryness to afford analytically pure product (NMR). Spectral data matches literature values.\textsuperscript{1}

\textit{General polymerization catalyst preparation:} To an oven dried Schlenk flask equipped with magnetic stir bar was added ((\textit{S})-BINOLate)Ti(O\textsuperscript{i}Pr)\textsubscript{2} (332 mg, 0.737 mmol, 1.0 equiv). To the solids was added AlEt\textsubscript{3} (1M solution in hexane, 2.95 mL, 2.95 mmol, 4.0 equiv) with stirring. The solution turned immediately dark brown with mild exotherm. The solution was then stirred at room temperature to ensure complete dissolution and then hexane removed completely under vacuum. The resulting dark brown viscous oil was then diluted with 7.25 mL of cumene (0.1 M) and the resulting solution stirred for 1 h at room temperature and 1 h at 150 ℃. The dark brown solution was then used immediately or stored in a -30 ℃ freezer in a glove box in a sealed vial. The catalyst solution retains activity for at least 2 months if properly handled under inert atmosphere. 

\textit{Polymerization Procedure:} In a glove box with a custom-built deposition apparatus suitable substrates (\~1 cm\textsuperscript{2} glass, quartz, sapphire, or silicon wafers) were placed within the 250 mL Schlenk flask deposition chamber equipped with separate vacuum and acetylene gas lines. The previously synthesized catalyst solution was diluted to 0.001-0.03 M with toluene for desired catalyst activity and film thickness. The dilute catalyst solution was then dropped onto the substrate to cover the surface via syringe with 22G needle, apparatus sealed, evacuated, and refilled with 1 atm of acetylene gas. Polymerization is observable immediately as dark black/purple fibrils than form a mass. At desired thickness, polymerization was ceased by evacuation of the chamber, and evacuation continued until the films were visibly dry. The deposited substrate was removed from the chamber and immediately washed thoroughly with toluene as a gentle stream from a pipet. The washed substrate was then thoroughly dried under vacuum and subjected to analysis. 

\section{Experimental Spectroscopy}

The setup for measuring the absorption and CD spectra is shown in Figure S3 where a monochromator equipped with a Xenon/tungsten lamp (as per required wavelength) sends an incident light beam that passes through the linear polarizer (45\textsuperscript{o}), photo elastic modulator (PEM), and to the sample, and the transmitted light intensity is measured using a silicon detector equipped with pre-amplifier.

 The absorbance spectrum of the regular and chiral polyacetylene was calculated from the measured optical transmission (\textit{T}) spectrum as

 \begin{equation}
     A = -\log(\frac{\Delta T}{T_0}) =-\ln\bigg( \frac{\frac{\Delta T}{T_0}}{\ln(10)}\bigg).
 \end{equation}

 The difference in the absorbance ($\Delta A = A_L - A_R$) of the two light helicities, namely left- and right-circularly polarized light, is given by

 \begin{equation}
     \Delta A \approx \frac{\Delta T}{T ln(10)},  
 \end{equation}
where $\Delta T = T_L - T_R$ the difference between the transmission $T_L$ of left-circularly polarized light and $T_R$ right circularly polarized light.

Subsequently, the circular dichroism (CD) spectrum of the film was calculated using the relation
\begin{equation}
    CD = \Delta A *( \frac{ln(10)}{4} ) * ( \frac{1.8\times10^5}{\pi} ),
\end{equation}
From Equations (S4) and (S5), we may express the CD value in mdeg as: 
\begin{equation}
    CD\approx -\frac{\Delta T}{T ln(10)}  *32980.
\end{equation}

The transmission $T$ was measured by the first lock-in amplifier, which was referenced to the chopper operating at 210 Hz. The differential transmission, $\Delta T$ , was measured by the second lock-in amplifier, which was referenced to the photoelastic modulator (\textbf{PEM} 200 from Hinds Instruments) operating at 50 kHz. In Equation (S4), $\Delta T=2X_2$, where $X_2$ is the output of the second lock-in that operates in phase with the PEM, and $T=X_1$ where $X_1$ is the output of the first lock-in operating in phase with the chopper. The factor of two arises from the chopper, which reduces the effective integration time of the second lock-in by half. The phase of the second lock-in was calibrated by replacing the sample with a quarter-wave plate and a linear polarizer to simulate pure circular polarization. 

\section{Theoretical Methods and Computational Details}
 
\textit{Theoretical Methods:} Ground-state geometry and electronic structure were found from the self-consistent DFT equations which minimize the total energies. Excited electronic states were found using time-dependent density functional theory (TDDFT) by solving the Casida equations. Briefly, solving the random phase approximaton (RPA),
\begin{equation}
    \begin{bmatrix}
        \textbf{A}&\textbf{B}\\
        -\textbf{B}&-\textbf{A}
    \end{bmatrix}
    \begin{bmatrix}
        \textbf{X}^{0\alpha}\\
        \textbf{Y}^{0\alpha}
    \end{bmatrix} =
    E^{0\alpha}
    \begin{bmatrix}
        \textbf{X}^{0\alpha}\\
        \textbf{Y}^{0\alpha}
    \end{bmatrix},
\end{equation}
one obtains the transition density matrix
\begin{equation}
    \boldsymbol{\xi}^{0\alpha} = 
    \begin{bmatrix}
    \boldsymbol{0}& \boldsymbol{X}^{0\alpha}\\
    \boldsymbol{Y}^{0\alpha}& \boldsymbol{0}
    \end{bmatrix},~
    {\boldsymbol{\xi}^{0\alpha}}^\dagger = 
    \begin{bmatrix}
    \boldsymbol{0}& \boldsymbol{Y}^{0\alpha}\\
    \boldsymbol{X}^{0\alpha}& \boldsymbol{0}
    \end{bmatrix},
\end{equation}
with normalization $\mathrm{Tr}[\boldsymbol{\xi}^{0\alpha}(\boldsymbol{O}-\boldsymbol{V}){\boldsymbol{\xi}^{0\alpha}}^\dagger] = \textbf{X}^{0\alpha}\cdot \textbf{X}^{0\alpha} - \textbf{Y}^{0\alpha}\cdot \textbf{Y}^{0\alpha}$, where $\boldsymbol{O}$ ($\boldsymbol{V}$) is the identity matrix for the occupied (virtual) MO subspace. From the transition electric dipole $\vec{d}^{0\alpha}$ and transition magnetic dipole  $\vec{m}^{0\alpha}$ expressed as

\begin{equation}
    \vec{d}^{0\alpha} = -\langle 0 |  \vec{\hat{r}} | \alpha \rangle,
\end{equation}
\begin{equation}
    \vec{m}^{\alpha 0} = \langle 0 |  \vec{\hat{r}} \times \vec{\hat{p}} | \alpha \rangle,
\end{equation}

we can compute the oscillator strength $f^{0\alpha}$ isotropically averaged optical rotation represented by the rotatory strength $R^{0\alpha}$ expressed as
\begin{equation}
    f^{0\alpha}=\frac{2}{3}E^{o\alpha}|\vec{d}^{0\alpha}|^2,
\end{equation}
\begin{equation}
    R^{0\alpha} = Im(\vec{d}^{0\alpha} \cdot \vec{m}^{\alpha 0}).
\end{equation}

The extinction coefficient,$\epsilon$, and differential extinction coefficient,$\Delta \epsilon$, describing circular dichroism (CD), is then generated with equations 

\begin{equation}
    \epsilon=\frac{1.31\times10^5}{\sigma} \sum_{\alpha} f^{o\alpha}\delta(E - E^{o\alpha} ),
\end{equation}

\begin{equation}
    \Delta \epsilon=\frac{1}{22.97\times10^{-40}\times\sigma} \sum_{\alpha} E^{o\alpha}R^{o\alpha}\delta(E - E^{o\alpha} ),
\end{equation}

 where the delta function is broadened as a Gaussian distribution. Note that oscillator strength  is dimensionless and rotatory strengths  are in cgs units of $10^{-40} \frac{erg \space esu}{G\space cm}$.

 To compare the relative intensities of the linear and polarized absorption, we computed the anisotropy factor from the computed extinction and differential extinction coefficients $g=\frac{\Delta \epsilon}{\epsilon}$. 

 \textit{Computational Details:} Ground- and excited-state calculations were performed using DFT and TD-DFT, respectively, as implemented in the \textit{Gaussian 16} package.\cite{g16} All simulations employed the CAM-B3LYP exchange-correlation functional\cite{YANAI2004}, the 6-31G* basis set\cite{Collins1976}, and a polarizable continuum model \cite{Klamt1993,Barone1998} with a dielectric constant of 6 to approximate the polyacetylene film environment.

The ground-state structure of achiral \textit{cis }polyacetylene modeled as a neutral singlet C\textsubscript{100}H\textsubscript{102} chain, constructed by repeating 25 C$_4$H$_4$ units and terminating edge carbons with two hydrogens. All oligomer geometries were fully optimized prior to analysis. From the achiral oligomer we next induce helicity into the chain by rotating the dihedral angle between adjacent monomers along the chain by an angle $\theta=\frac{360}{(N-2)t}$ where$N$ is the number of monomers and $t$ is the number of turns (related to pitch). We explore models with turn number $t\in {(1/2,1,3/2,2,5/2,3})$  The \textit{R}- and \textit{S}- helical enantiomers are generated by applying positive and negative rotations, respectively.  

\section{Natural Atomic Orbital Analysis}

For the analysis of nearest-neighbor hopping parameters $t_{ij} = \braket{\phi_i|F|\phi_j}$ we compute a localized basis-set in the Natural Atomic Orbital (NAO) framework.\cite{Reed1985} In polyacetylene the electronic and optical properties are determined by the $\sigma$ and $\pi$  bonds formed by the overlap of the $2p_{x,y,z}$ atomic basis functions. For planar (CH)$_x$ with $z$ as the out-of-plane direction the $2p_z$ orbitals on adjacent Carbon sites form the $\pi$ bond while the in-plane $2p_x$ and $2p_y$ orbitals participate in $sp_2$ hybridization. When torsion is introduced the out-of-plane direction of the $2p$ orbital participating in the $\pi$ bond will vary along the chain. We define the orbital participating in the $\pi$ bond from the $i^{th}$ Carbon ion as a linear combination of the NAO $2p$ basis 

\begin{equation}
    \ket{\phi_i} =n_x\ket{p_x^i} + n_y\ket{p_y^i}+n_z\ket{p_z^i} ,
\end{equation}
where $\vec{n}=(n_x,n_y,n_z)$ is the local out-of-plane axis of the \textbf{C-C} bond determined as 
\begin{equation}
    \vec{n}=(\frac{N_x}{|N|},\frac{N_y}{|N|},\frac{N_z}{|N|} ),
\end{equation}
\begin{equation}
    \vec{N}=\vec{v}_{i+1} \times\vec{v}_i, 
\end{equation}
\begin{equation}
    \vec{v}_i=\vec{r}_{i} -\vec{r}_{i-1} ,
\end{equation}
with $\vec{N}$ being the normal vector, $\vec{v}_i$ being the bond direction vector between adjacent \textbf{C-C} ions in the bond, and $\vec{r}_i$ being the Cartesian coordinates of the $i^{th}$ ion.

\section{Exciton and Transition Dipole Symmetry}
\begin{figure}[!h]
    \centering
    \includegraphics[width=\linewidth]{PA_spectra_sticks_TD.jpg}
    \caption{\footnotesize
    (a) Absorption and (b) Circular Dichroism spectra of the \textit{helical }polyacetylene structures wrapped at 0.00$\degree$, 3.75$\degree$, 7.50$\degree$, and 15.00$\degree$. (c) Transition density matrix $\boldsymbol{Q}_{AB}$ for the first five excitons for 0.00$\degree$ and 15.00$\degree$ as well as the transition chiral tensor $\boldsymbol{R}_{AB}$. In all cases, the matrices are summed onto each atom as $Q_{AB} = \sum_{\mu \in A} \sum_{\nu \in B} |Q_{\mu\nu}|^2$ and $R_{AB} = \sum_{\mu \in A} \sum_{\nu \in B} R_{\mu\nu}$, respectively.
    }
    \label{FIG_SUPP:SPECTRA_STICKS_HELIX_TDM_TCT}
\end{figure}

\begin{figure}[!h]
    \centering
    \includegraphics[width=\linewidth]{PA_spectra_sticks_TD_twist.jpg}
    \caption{\footnotesize
    (a) Absorption and (b) Circular Dichroism spectra of the \textit{twisted} polyacetylene structures wrapped at 0.00$\degree$, 3.75$\degree$, 7.50$\degree$, and 15.00$\degree$. (c) Transition density matrix $\boldsymbol{Q}_{AB}$ for the first five excitons for 0.00$\degree$ and 15.00$\degree$ as well as the transition chiral tensor $\boldsymbol{R}_{AB}$. In all cases, the matrices are summed onto each atom as $Q_{AB} = \sum_{\mu \in A} \sum_{\nu \in B} |Q_{\mu\nu}|^2$ and $R_{AB} = \sum_{\mu \in A} \sum_{\nu \in B} R_{\mu\nu}$, respectively.
    }
    \label{FIG_SUPP:SPECTRA_STICKS_TWIST_TDM_TCT}
\end{figure}


\clearpage

{\small
\bibliography{main.bib}
}